# LATTICE QCD[*]

Aida X. El-Khadra

*Physics Department, Ohio State University, 174 W 18th Ave, Columbus, OH 43210*

## Abstract

The status of Lattice QCD is reviewed with respect to results that are relevant to Standard Model phenomenology. I argue that in a few simple cases all (or almost all) systematic errors from the lattice calculation are under control. Lattice QCD calculations of quarkonia ($b\bar{b}$ and $c\bar{c}$) have yielded precise determinations of the strong coupling, $\alpha_s$, and the heavy quark masses. Recent work on the kaon $B$ parameter has also led to a significant reduction in systematic errors, resulting in a precise determination of this hadronic matrix element. I discuss recent progress in other calculations of potential phenomenological significance, as well as problems and prospects.

---

# 1 Introduction and Motivation

Since this appears to be the first review of lattice QCD in this series, I would like to give a general perspective before I start the discussion of this specific review.

By now, we have accumulated a large body of circumstantial evidence that QCD is the correct theory of the strong interactions. What is sorely missing is a first-principles understanding of the non-perturbative effects QCD gives rise to, the most dramatic of them being the observed hadron spectrum. On the practical side, this lack of understanding limits the extraction of Standard Model parameters from experimental measurements. Lattice field theory offers a systematic first principles approach to solving QCD. Calculations in lattice QCD can be performed using Monte Carlo methods.

Beyond QCD, lattice field theory has been used to study many other interesting topics, such as electroweak symmetry breaking, chiral gauge theories, quantum gravity, and statistical systems, none of which I will discuss in this review.

I will limit the discussion in this review to calculations directly relevant to the extraction of Standard Model parameters. Needless to say, the field of lattice QCD is much richer than my limited discussion would indicate.

Most of the work of phenomenological relevance is done in what is generally referred to as the "quenched" (and sometimes as the "valence") approximation. In this approximation (see section 2.1) gluons are not allowed to split into quark - anti-quark pairs (sea quarks). The obvious question from the non-expert is why such an arbitrary approximation is used at all in lattice QCD calculations, if the purported goal is a first-principles understanding of QCD ? The answer comes in two parts. Perturbative arguments suggest that most of the effect of sea quarks can be absorbed into the coupling constant. Of course, non-perturbative effects need not be so benign, and there certainly are physical quantities for which they are not. However, recent calculations of the light hadron spectrum (see section 4) indicate that after taking the continuum limit, quenched QCD can give us an acceptable phenomenology. The second part of the answer is calculational convenience. Calculations in the quenched approximation are simple enough to allow the study and control of other systematic errors that arise in lattice QCD calculations as detailed in section 2. They include, but are not limited to, finite lattice spacing, finite volume, and the extrapolation to physical light-quark masses.

Algorithmic progress in reducing the computational expense of including sea quarks together with the obvious increases in computing power (of which a significant part has been dedicated to this problem) is producing a few calculations that partially include sea quarks, with some phenomenological impact, as discussed in sections 3 and 5.1.

Two limits of QCD exist where the theory becomes simple enough to be tackled with analytic methods. These are the chiral and the heavy-quark limits. Both limits can be studied in lattice QCD. In particular, the effect of sea quarks can be analyzed using techniques developed in the continuum: chiral perturbation theory and non-relativistic potential models.

I would like to emphasize that lattice QCD is not a model. With all systematic errors under control, a lattice QCD result becomes a prediction of QCD.

The rest of this review proceeds as follows: Section 2 gives a brief introduction to lattice QCD methods without going into technical details, since a number of pedagogical introductions and reviews already exist in the literature [1]. Attention is focused on the sources of systematic errors that enter lattice QCD calculations with the aim to enable the non-expert to judge the reliability of a lattice QCD result of interest to her/him. The following sections 3-5 discuss physics results by topic. Section 6 finally concludes with some remarks about future prospects.

# 2  A Lattice QCD "Primer"

Starting with the Feynman path integral formulation in Euclidean space, the discretization of space-time (with lattice spacing $a$) regulates the integral at short distances or in the ultraviolet. A finite volume (of length $L$) is necessary for numerical techniques and also introduces an infrared cut-off or momentum-space discretization. The vacuum expectation of a Greens function, $\mathcal{G}$, which is a product of gauge and fermion fields, is defined as:

$$\langle \mathcal{G} \rangle = \lim_{L \to \infty} \lim_{a \to 0} \langle \mathcal{G} \rangle_{L,a} \; , \quad \langle \mathcal{G} \rangle_{L,a} = Z_{L,a}^{-1} \int \mathcal{D}\psi \mathcal{D}\bar{\psi} \mathcal{D}U \, \mathcal{G} \, e^{-S} \; . \tag{1}$$

$Z_{L,a}$ normalizes the expectation value. I have omitted spin and color indices for compactness. The gauge degrees of freedom are written as (path ordered) exponentials of the gauge field, $A_\mu$:

$$U_\mu(x) = e^{i \int_x^{x+a} dx' A_\mu(x')} \simeq e^{ia A_\mu(x)} \; , \tag{2}$$

which makes it easy to maintain gauge invariance. The link fields, $U$, are $SU(3)$ matrices. The (Euclidean) QCD action,

$$S = S_g + S_f \; , \quad S_g = \frac{1}{4g^2} \int d^4x \, F_{\mu\nu} F^{\mu\nu} \; , \quad S_f = \int d^4x \, \bar{\psi}(x)(\slashed{D} + m)\psi(x) \; . \tag{3}$$

is discretized, such that Eq. (3) is recovered in the the continuum ($a \to 0$) limit:

$$S_{\text{lat}} = S + \mathcal{O}(a^n) \; , \quad n \geq 1 \; . \tag{4}$$

I will not go into the explicit formulations of $S_{\text{lat}}$ here, but instead refer the reader to Ref. [1]. The most common form for the gauge action is Wilson's [2], written in terms of plaquettes – products of $U$ fields around the smallest closed loop on a lattice. For fermions the situation is more complicated. The discretization of

$$M \equiv \slashed{D} + m \; , \tag{5}$$

is a sparse, finite dimensional matrix. Two different approaches are in use. In Wilson's formulation [3] chiral invariance is explicitly broken, but restored in the continuum limit. The pay-off is a solution of the so-called fermion doubling problem. Staggered fermions [4] keep a $U(1)$ chiral symmetry at the expense of dealing with 4 degenerate flavors of fermions.

Eq. (1) emphasizes that QCD is a limit of lattice QCD. However, in numerical calculations these limits cannot be taken explicitly, only by extrapolation. This is feasible, because theoretical guidance for both limits is available. The zero-lattice-spacing limit is guided by asymptotic freedom, since the lattice spacing is related to the gauge coupling by the renormalization group. Quantum field theories in large but finite volumes have also been analyzed theoretically [5].

In a numerical calculation the limits are taken by considering a series of lattices, as illustrated in Figure 1. While keeping the physical volume (or $L$) fixed, the lattice spacing is successively reduced; then, keeping the lattice spacing fixed the volume is increased. The calculation is in the continuum (infinite volume) limit once the hadron spectrum or matrix elements of interest become independent of the lattice spacing (volume).

In practice, however, limitations in computational resources do not permit the ideal lattice QCD calculation just described. In particular, the computational cost of reducing the lattice spacing naively scales like $(L/a)^4$. (The computational cost is really higher, because of numerical problems at smaller lattice spacings.) Eq. (4) illustrates an alternative. By improving the discretization errors in the lattice action (and operators), the continuum limit can be reached at coarser lattice spacings than before. Simulations with improved actions can come at only a slightly higher computational price. The ideas underlying improvement were developed some time ago [6, 7, 8], and have since been revitalized [9, 10, 11, 12].

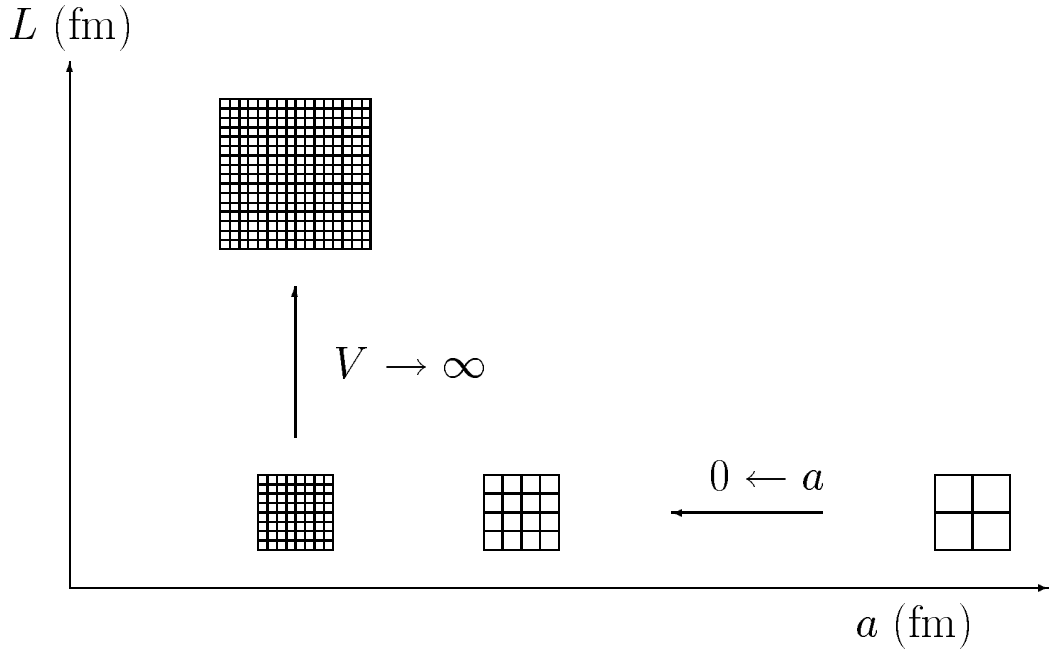

Figure 1: Illustration of the continuum and infinite-volume limits.

If the quark mass is large compared to the typical QCD scale, $\Lambda_{QCD}$, effective theories are most adequate in describing the physics [13]. In that case, the lattice spacing cannot be taken to zero. Lattice-Spacing errors can, however, be systematically reduced by improvement [14].

Typical lattice QCD calculations have lattice spacings in the range $0.05\,\mathrm{fm} \lesssim a \lesssim 0.2\mathrm{fm}$, and finite volumes with spatial lengths in the range $1.0\,\mathrm{fm} \lesssim L \lesssim 3.0$ fm.

Now that the problem is (more or less) set up, I will briefly go through the organization of a typical lattice QCD calculation in the next 3 subsections.

## 2.1 Gauge Configurations

The Euclidean path integral has a formal analogy to statistical mechanics. This analogy has been exploited in adapting tools, originally developed for statistical mechanics systems, to lattice QCD.

Because the fermion fields anti-commute (they belong to a Grassman Algebra), their integration can be done analytically, leaving only the integral over the gauge fields to be done numerically. For example, the normalization $Z$, becomes

$$Z_{L,a} = \int \mathcal{D}U \, \det M[U] \, e^{-S_g[U]} \quad . \tag{6}$$

The only viable numerical technique for the evaluation of such a high-dimensional integral is the Monte Carlo method. New gauge fields are sampled with probability $\mathcal{D}U \det M[U] e^{-S_g[U]}$ using importance sampling. A gauge configuration is a lattice ensemble of such gauge fields. Approximating the integral over gauge fields in Eq. (1) with a finite sum over the gauge configurations introduces a statistical error.

Rather than going into details about algorithms, let me just note that the presence of the determinant, $\det M$, complicates the process of choosing new gauge configurations. The quenched approximation [15] sets $\det M = 1$ in Eq. (6), which reduces the computational effort by several orders of magnitude at the expense of neglecting vacuum polarization. In perturbation theory this can be compensated by a shift in the coupling constant. The interaction of valence quarks with gluons is, however, still treated correctly.

## 2.2 Quark Propagators

Calculating a quark propagator

$$\langle \psi(x)\bar{\psi}(y) \rangle_{L,a} = Z_{L,a}^{-1} \int \mathcal{D}U \det M[U] \, M_{xy}^{-1}[U] \, e^{-S_g[U]} \tag{7}$$

reduces to the computational problem of calculating the inverse of the matrix $M$. $M$ is singular for massless quarks, and the inversion becomes difficult (if not impossible) for small quark masses, depending on the volume, lattice spacing and sophistication of the algorithm used. In practice this has restricted calculations involving light quarks to masses $m_q \gtrsim m_s/3$, making extrapolations to the physical $m_{u,d}$ quark masses necessary. In some cases, this extrapolation can be guided by chiral perturbation theory. However, if the quark-mass dependence is not known a priori, control of this systematic error is possible only if the observed mass dependence is weak.

## 2.3 Hadron Propagators

Hadron masses are usually extracted [15] from 2 - point functions of the form:

$$\langle G_\pi(x,y) \rangle = \langle \chi_\pi(x) \chi_\pi^\dagger(y) \rangle \ , \tag{8}$$

where the interpolating field, $\chi(x)$, has the quantum numbers of the hadron in question. A simple choice for, say, a pion field is $\chi_\pi(x) = \bar{\psi}(x)\gamma_5\psi(x)$. It is easily shown that the hadron propagator in Eq. (8) is a product of quark propagators in the background of the gauge fields:

$$\langle G_\pi(x,y) \rangle = Z_{L,a}^{-1} \int \mathcal{D}U \det M[U] \, M_{yx}^{-1}[U] \, \gamma^5 \, M_{xy}^{-1}[U] \, \gamma^5 \, e^{-S_g[U]} \tag{9}$$

Projecting out a specific 3 - momentum, **p**, it can be shown that the hadron 2 - point function is a sum over all states (labeled $n$) with the same quantum numbers as, in this case, the pion:

$$\begin{aligned} G_\pi(\mathbf{p}, t) &= \sum_{\mathbf{x}} e^{i\mathbf{p}\cdot\mathbf{x}} \langle G_\pi(x, 0) \rangle \\ &= \sum_n C_n^2 e^{-E_n t} \ , \quad C_n = \langle 0|\chi_\pi|n \rangle \ . \end{aligned} \tag{10}$$

The lowest exponential, $E_1 \equiv E_\pi$, dominates at large Euclidean times, allowing the extraction of the pion mass from the exponential fall-off of $G_\pi(\mathbf{p} = 0, t)$ at large times.

The coefficients, $C_n$, describe the overlap of the operators $\chi$ with the different states $n$. The presence of excited states in Eq. (10) introduces a systematic error into the hadron mass determination, which has to be controlled. A smart choice of $\chi$ minimizes the contributions of unwanted excited states to the sum in eq. (10), and thus allows for better control of this systematic error and the extraction of the ground-state mass at small time slices. For some mesons, the statistical errors (from the Monte Carlo integration) grow with Euclidean time, which makes the ground-state extraction difficult. Examples include mesons with finite momentum, orbitally excited states, and mesons with one infinitely heavy (static) quark. In such cases, it becomes necessary to carefully project out excited state contributions by constructing a matrix of hadron operators. Such techniques can also yield radially excited state masses.

The coefficients $C_n$ become matrix elements if one of the operators $\chi$ is substituted with a current. Taking, for example, the axial current, $A_\mu(x) = \bar\psi(x)\gamma_\mu\gamma_5\psi(x)$, would leave $\langle 0|A_\mu|\pi\rangle$ in the ground-state contribution in Eq. (10), allowing the extraction of the pion decay constant.

Three - point functions are needed in order to calculate matrix elements that are relevant for $K - \bar K$ mixing, semi-leptonic decays, etc., in a straightforward generalization of the above discussion. Let us consider the transition $B \to A$, governed by $\langle A|\mathcal{J}|B\rangle$:

$$G_{A\mathcal{J}B}(\mathbf{p}_A, \mathbf{p}_\mathcal{J}, t_A, t_\mathcal{J}) = \sum_{\mathbf{x}_A, \mathbf{x}_B} e^{i\mathbf{p}_A\cdot\mathbf{x}_A} e^{i\mathbf{p}_\mathcal{J}\cdot\mathbf{x}_\mathcal{J}} \langle \chi_A(x_A)\mathcal{J}(x_\mathcal{J})\chi_B^\dagger(0) \rangle \quad (11)$$

$$\stackrel{t_A, t_B \to \infty}{\longrightarrow} e^{-E_A(t_A - t_\mathcal{J})} e^{-E_B t_\mathcal{J}} \langle 0|\chi_A|A\rangle \underline{\langle A|\mathcal{J}|B\rangle} \langle B|\chi_B^\dagger|0\rangle \quad (12)$$

Efficient techniques [16] have been developed to deal with the more complicated products of quark propagators that appear in the corresponding path integral (analogous to Eq. (9)). The calculational effort is about twice of that necessary for hadron masses and decay constants per momentum combination $\mathbf{p}_A$, $\mathbf{p}_\mathcal{J}$.

## 2.4 Perturbation Theory

Lattice QCD calculations use perturbation theory in several places:

- It guides the approach to the continuum limit.

- Short-Distance quantities can be calculated non-perturbatively and compared to their perturbative expansions. Ref. [17] showed that indeed, 1-loop perturbation theory describes most quantities considered to $3 - 5\%$, if a renormalized coupling like $\alpha_{\overline{MS}}$ (rather than the bare lattice coupling) is used and the lattice spacing $a \lesssim 0.2$ fm.

- Matrix elements calculated with a lattice regulator have to be matched to their continuum counterpart by perturbation theory.

- Because quarks are confined into hadrons, quark masses are always scheme dependent. Perturbation theory is used to convert non perturbatively determined lattice quark masses to the perturbative continuum masses such as the pole or $\overline{MS}$ masses. Similarly, the gauge coupling can be determined non-perturbatively using lattice QCD and converted to the $\overline{MS}$ scheme at large momenta (see also section 3).

The lattice regulator breaks Lorentz (or Euclidean) invariance, which complicates perturbative calculations relative to those performed with Lorentz (or Euclidean) invariant regulators, such as dimensional regularization. This has prompted the development of computational techniques for higher loop perturbative calculations [18]. (Numerical) techniques for non-perturbative calculations of renormalization constants have also been developed [19, 20]. Such techniques are very promising, because every time a new action or new operators are considered, not only must the programs be changed but also the perturbation theory has to be redone.

# 3  Quarkonium Spectroscopy

Quarkonia, mesons containing a heavy quark and anti-quark, are at present the best understood hadronic systems. Both, the charm and bottom quark masses are large compared to the typical QCD scale, $\Lambda_{QCD}$. The $b\bar{b}$ and $c\bar{c}$ bound states are therefore governed by non-relativistic dynamics. While the QCD potential was not known from first principles, relatively simple guesses for phenomenological potentials have proven quite successful in describing the experimentally measured bound state spectra of quarkonia [21].

As has been argued by Lepage [22], quarkonia are also the easiest systems to study with lattice QCD, and systematic errors can be analyzed using potential models. For example, finite-volume errors are much easier to control for quarkonia than for light hadrons; they are negligible if volumes with $L \gtrsim 1.5$ fm are used. Lattice-Spacing errors, on the other hand, can be larger for quarkonia and need to be considered. However, the size of lattice-spacing errors in a numerical simulation of quarkonia can be *anticipated* by calculating expectation values of the corresponding operators using potential model wave functions. Control over systematic errors in turn allows the extraction of Standard Model parameters from the quarkonia spectra.

Two formulations have been used in calculations of these spectra. In the non-relativistic limit the QCD action can written as an expansion in powers of $v^2$ (or $1/m$), where $v$ is the velocity of the heavy quark inside the boundstate [13], which I henceforth shall refer to as NRQCD. Lepage and collaborators [14] have adapted this formalism to the lattice regulator. Two groups have performed numerical calculations of quarkonia in this approach. In refs. [23, 24] the NRQCD action is used to calculate the $b\bar{b}$ and $c\bar{c}$ spectra, including terms up to $\mathcal{O}(mv^4)$ and $\mathcal{O}(a^2)$. In addition to calculations in the quenched approximation, this group is also using gauge configurations that include 2 flavors of sea quarks with mass $m_q \sim \frac{1}{2}m_s$ to calculate the $b\bar{b}$ spectrum [25, 26]. The leading order NRQCD action is used in Ref. [27] for a calculation of the $b\bar{b}$ spectrum in the quenched approximation.

The Fermilab group [10] developed a generalization of previous approaches, which encompasses the non-relativistic limit for heavy quarks as well as Wilson's relativistic action for light quarks. Lattice-Spacing artifacts are analyzed for quarks with arbitrary mass.

Ref. [28] uses this approach to calculate the $b\bar{b}$ and $c\bar{c}$ spectra in the quenched approximation. They considered the effect of reducing lattice-spacing errors from $\mathcal{O}(a)$ to $\mathcal{O}(a^2)$. The results for the $b\bar{b}$ and $c\bar{c}$ spectra from all groups are summarized in Figures 2 and 3.

The agreement between the experimentally-observed spectrum and lattice QCD calculations is impressive. As indicated in the preceding paragraphs, the lattice artifacts are different for all groups. Figures 2 and 3 therefore emphasize the level of control over systematic errors.

All groups use gauge configurations generated with the Wilson action leaving $\mathcal{O}(a^2)$ lattice-spacing errors in the results. Since the heavy-quark actions have been [23], or can be improved through $\mathcal{O}(a^2)$, an improved gauge action [6] should also be used, effectively reducing lattice-spacing errors to a negligible level.

The first results with 2 flavors of degenerate sea quarks have appeared [26, 29] with lattice-spacing and finite-volume errors similar to the quenched calculations, significantly reducing this systematic error. However, several systematic effects associated with the inclusion of sea quarks still have to be studied. They include the dependence on the quarkonium spectrum of the number of flavors of sea quarks and the sea-quark action (staggered vs. Wilson). The inclusion of sea quarks with realistic light-quark masses is very difficult. However, quarkonia are expected to depend only very mildly on the masses of the light quarks. This systematic error has not been included yet and should be checked numerically.

Quarkonia were, upon their discovery, called the hydrogen atoms of particle physics. Their non-relativistic nature justified the use of potential models, which gave a nice, phenomenological understanding of these systems. This phenomenology is at present useful to control systematic errors in lattice QCD calculations of $b\bar{b}$ and $c\bar{c}$ spectra. However, we are quickly moving towards truly first-principles calculations of quarkonia using lattice QCD, thereby testing QCD non-perturbatively. In this sense, quarkonia are still the hydrogen atoms of particle physics.

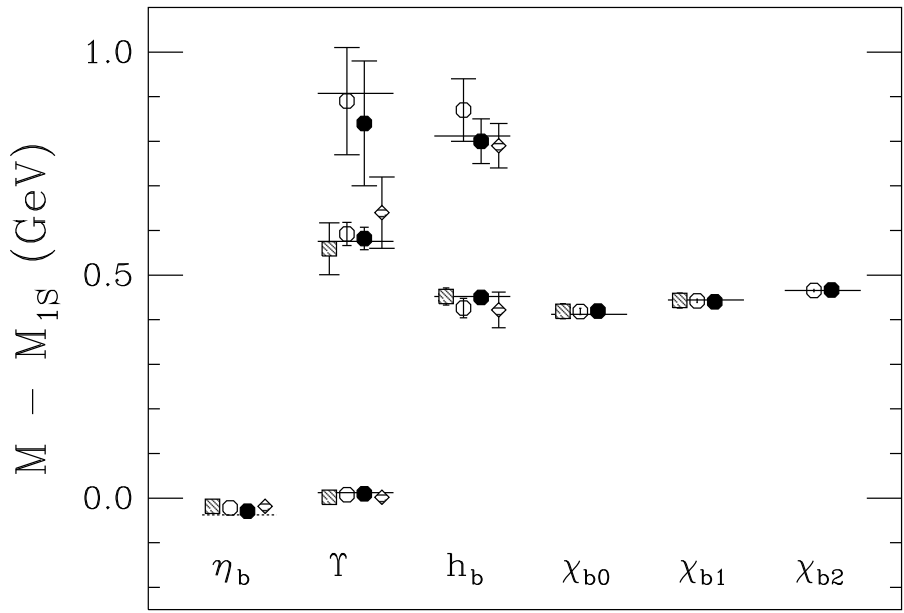

Figure 2: A comparison of lattice QCD results for the $b\bar{b}$ spectrum (statistical errors only). -: Experiment; □: FNAL [28]; ○: NRQCD ($n_f = 0$) [23]; •: NRQCD ($n_f = 2$) [25]; ◇: UK(NR)QCD [27].

Precise determinations of the Standard Model parameters $\alpha_s$, $m_b$, $m_c$, are by-products of this work. They are discussed in the following two subsections.

## 3.1 The Strong Coupling from Quarkonia

At present, the QCD coupling, $\alpha_s$, is determined from many different experiments, performed at energies ranging from a few to 90 GeV [30]. In most cases perturbation theory is used to extract $\alpha_s$ from the experimental information. Experimental and theoretical progress over the last few years has made these determinations increasingly precise. However, all determinations, including those based on lattice QCD, rely on phenomenologically-estimated corrections and uncertainties from non-perturbative effects. These effects will eventually (or already do) limit the accuracy of the coupling constant determination. When lattice QCD is used the limiting uncertainty comes from the (total or partial) omission of sea quarks in numerical simulations. The determination of the strong coupling, $\alpha_s$, proceeds in three steps:

1. The experimental input to the strong coupling determination is a mass or mass splitting, from which by comparison with the corresponding lattice quantity the scale, $a^{-1}$, is determined in physical units. For this purpose, one should identify quantities that are insensitive to lattice errors. In quarkonia, spin-averaged splittings are good candidates. The experimentally observed 1P-1S and 2S-1S splittings depend only mildly on the quark mass (for masses between bottom and charm). Higher-Order lattice-spacing errors for these splittings are estimated in Refs. [26, 28] to be small. In Ref. [28] the calculation was performed at 3 different lattice spacings.

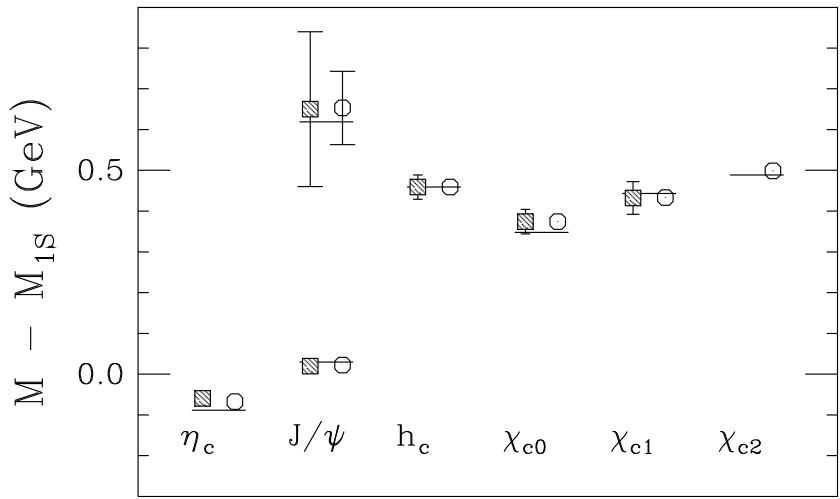

Figure 3: A comparison of lattice QCD results for the $c\bar{c}$ spectrum (statistical errors only). -: Experiment; □: FNAL [28]; ○: NRQCD ($n_f = 0$) [24]

2. Within the framework of lattice QCD the conversion from the bare to a renormalized coupling can, in principle, be made non-perturbatively. In the definition of a renormalized coupling, systematic uncertainties should be controllable, and at short distances, its (perturbative) relation to other conventional definitions calculable. For example, a renormalized coupling can be defined from the non-perturbatively computed heavy-quark potential [31]. In Ref. [32] a renormalized coupling is defined non-perturbatively through the Schrödinger functional. The authors compute the evolution of the coupling non-perturbatively using a finite size scaling technique, which allows them to vary the momentum scales by an order of magnitude. The strong coupling can also be computed from the three-gluon vertex, suitably defined on the lattice [33].

An alternative is to define a renormalized coupling through short distance lattice quantities, like small Wilson loops or Creutz ratios. For example, the coupling defined from the plaquette, $\alpha_P = -3\ln\langle\mathrm{Tr}\,U_P\rangle/4\pi$, can be expressed in terms of $\alpha_V$ (or $\alpha_{\overline{\mathrm{MS}}}$) by [17]:

$$\alpha_P = \alpha_V(q)[1 - 1.19\alpha_V(q) + \mathcal{O}(\alpha_V^2)] \qquad (13)$$

at $q = 3.41/a$, close to the ultraviolet cut-off. $\alpha_V$ is related to the more commonly used $\overline{\mathrm{MS}}$ coupling by

$$\alpha_{\overline{\mathrm{MS}}}(Q) = \alpha_V(e^{5/6}Q)(1 + \frac{2}{\pi}\alpha_V + \ldots) \quad . \qquad (14)$$

The size of higher-order corrections associated with the above defined coupling constants can be tested by comparing perturbative predictions for short-distance lattice quantities with non-perturbative results [17].

Ref. [28] takes a 5% uncertainty associated with perturbation theory from this analysis. The comparison of the non-perturbative coupling from Ref. [32] to perturbative predictions for this coupling using Eq. (13) is consistent with this error estimate.

In Ref. [26] the next-to-next-to-leading order corrections to Eq. (13) have been calculated numerically from the observed deviations (from 1-loop perturbation theory) in small Wilson loops and Creutz ratios (up to size 3) at several very small lattice spacings. The dominant perturbative error then comes from the conversion to the $\overline{\text{MS}}$ coupling, which is only known to 1-loop.

The relation of the plaquette coupling in Eq. (13) to the $\overline{\text{MS}}$ coupling will very soon be known to 2-loops [34], significantly reducing the uncertainty due to the use of perturbation theory.

3. Calculations that properly include all sea-quark effects do not yet exist. If we want to make contact with the "real world", these effects have to be estimated phenomenologically or extrapolated away.

The phenomenological correction necessary to account for the sea-quark effects omitted in calculations of quarkonia that use the quenched approximation have been discussed in detail in Refs. [35, 36]; it is the dominant systematic error in this calculation. Similar ideas were used to correct for sea-quark effects in early calculations of quarkonia spectra from the heavy-quark potential calculated in quenched lattice QCD [37].

By demanding that, say, the spin-averaged 1P-1S splitting calculated on the lattice reproduce the experimentally observed one (which sets the lattice spacing, $a^{-1}$, in physical units), the effective coupling of the quenched potential is in effect matched to the coupling of the effective 3 flavor potential at the typical momentum scale of the quarkonium states in question. The difference in the evolution of the zero flavor and 3,4 flavor couplings from the effective low-energy scale to the ultraviolet cut-off, where $\alpha_s$ is determined, is the perturbative estimate of the correction.

For comparison with other determinations of $\alpha_s$, the $\overline{\text{MS}}$ coupling can be evolved to the $Z$ mass scale. An average [30] of refs. [35, 36] yields for $\alpha_s$ from calculations in the quenched approximation:

$$\alpha_{\overline{\text{MS}}}^{(5)}(m_Z) = 0.110 \pm 0.006 \quad . \tag{15}$$

The phenomenological correction described in the previous paragraph has been tested from first principles in Ref. [29]. The 2-loop evolution of $n_f = 0$ and $n_f = 2$ $\overline{\text{MS}}$ couplings – extracted from calculations of the $c\bar{c}$ spectrum using the Wilson action in the quenched approximation and with 2 flavors of sea quarks respectively – to the low-energy scale gives consistent results. After correcting the 2 flavor result to $n_f = 3$ in the same manner as before and evolving $\alpha_{\overline{\text{MS}}}$ to the $Z$ mass, Ref. [29] finds

$$\alpha_{\overline{\text{MS}}}^{(5)}(m_Z) = 0.111 \pm 0.005 \tag{16}$$

in good agreement with the previous result in Eq. (15). The total error is now dominated by the rather large statistical errors and the perturbative uncertainty.

Ref. [26] used results for $\alpha_s$ from the $b\bar{b}$ spectrum with 0 and 2 flavors of sea quarks to extrapolate the inverse coupling to the physical 3 flavor case directly at the ultra violet momentum, $q = 3.41/a$. They obtain a result consistent with the old procedure, but with smaller errors:

$$\alpha_V^{(3)}(8.2\,\text{GeV}) = 0.196 \pm 0.003 \quad . \tag{17}$$

The error is dominated by the (small) statistical errors, not the extrapolation (in $n_f$) errors. The conversion to $\overline{\text{MS}}$ and evolution to the $Z$ mass then gives:

$$\alpha_{\overline{\text{MS}}}^{(5)}(m_Z) = 0.115 \pm 0.002 \quad , \tag{18}$$

with an error now dominated by the unknown higher orders in eq. (14).

The claimed result in Eq. (17) (or Eq. (18)) is the most accurate determination of the strong coupling constant to date. In order to confirm this result, it is desirable that the $b\bar{b}$ and $c\bar{c}$ spectra be calculated with heavy-quark actions based on [10] with the same level of statistical precision and care with respect to systematic errors as was done in Ref. [26]. Apart from this, the systematic errors associated with the inclusion of sea quarks into the simulation have to be checked, as outlined in section 3.

Phenomenological corrections are a necessary evil that enter most coupling constant determinations. In contrast, lattice QCD calculations with complete control over systematic errors will yield truly first-principles determinations of $\alpha_s$ from the experimentally observed hadron spectrum.

At present, determinations of $\alpha_s$ from quarkonia using lattice QCD are comparable in reliability and accuracy to other determinations based on perturbative QCD from high energy experiments. They are therefore part of the 1994 world average for $\alpha_s$ [30]. The phenomenological corrections for the most important sources of systematic errors in lattice QCD calculations of quarkonia are now being replaced by first-principles, which will significantly increase the accuracy of $\alpha_s$ determinations from quarkonia.

In a few years time, the world average for the strong coupling will be dominated by determinations of $\alpha_s$ using lattice QCD.

## 3.2 The Heavy Quark Masses

Because of confinement, the quark masses cannot be measured directly, but have to be inferred from experimental measurements of hadron masses, and depend on the calculational scheme employed.

In lattice QCD quark masses are determined non-perturbatively, by tuning the bare lattice quark mass ($m_Q^{\text{lat}}$) so that, for example, the experimentally observed $J/\psi$ (or $\Upsilon$) mass is reproduced by the calculation. For phenomenology useful quark masses are the perturbatively defined pole and $\overline{\text{MS}}$ masses, which the bare lattice mass can be related to by (1-loop) perturbation theory:

$$m_Q^{\text{pole}} = Z_m^{\text{pole}} m_Q^{\text{lat}} \quad , \qquad m_Q^{\overline{\text{MS}}}(m_Q) = Z_m^{\overline{\text{MS}}} m_Q^{\text{lat}} \quad . \tag{19}$$

The heavy-quark pole mass can be determined alternatively from a calculation of the binding energy, $E_{\text{bind}}$. The ground-state energy need not equal the mass of a non-relativistic system. The binding energy can be obtained by subtracting the perturbatively calculable heavy-quark rest energy from the ground-state energy. The pole mass is then:

$$m_Q^{\text{pole}} = \frac{1}{2}(M_{Q\bar{Q}}^{\text{exp}} - E_{\text{bind}}) \tag{20}$$

This method is insensitive to errors in tuning the bare mass, because the binding energy depends only mildly on the quark mass.

Of course, as always, all systematic errors arising from the lattice QCD calculation need to be under control for a phenomenologically interesting result; in particular, the systematic error introduced by the (partial) omission of sea quarks has to be removed. The short-distance corrections that introduced the dominant uncertainty to the $\alpha_s$ determination from quarkonia are absent for the pole mass determination, because this effective mass does not run for momenta below it's mass.

Ref. [25] used both methods described above for a determination of the $b$ quark pole mass from a lattice QCD calculation of the $b\bar{b}$ spectrum. As expected, a comparison of their results with zero and 2 flavors of sea quarks finds compatible results for the pole mass:

$$m_b^{\text{pole}} = (5.0 \pm 0.2)\text{ GeV} \tag{21}$$

For the $\overline{\text{MS}}$ mass, Ref. [25] quotes $m_b^{\overline{\text{MS}}}(m_b) = 4.0(1)$ GeV. The error in both results is dominated by perturbation theory.

A similar analysis is being performed in Ref. [38] for the charm quark mass from the charmonium spectrum. A preliminary result is $m_c^{\text{pole}} = 1.5(2)$ GeV.

The $\overline{\text{MS}}$ mass for the charm quark has also been determined from a compilation of $D$ meson calculations in the quenched approximation [39], with $m_c^{\overline{\text{MS}}}(2\,\text{GeV}) = 1.47(28)$ GeV. The error includes statistical errors from the original calculations and the perturbative error. However sea-quark effects cannot, in this case, be estimated phenomenologically, leaving this systematic error uncontrolled.

# 4 Light Hadron Spectrum

The calculation of, say, the proton mass from first principles is one of the original problems lattice methods hoped to solve. Unfortunately, the systematic errors in the lattice calculations do not yet allow for first-principles calculations of the light hadron spectrum. Consequently, while a first-principles understanding of the (low lying) $b\bar{b}$ and $c\bar{c}$ spectra lies just around the corner, as I argued in the previous section, the light hadron spectrum is one of the hardest problems, most likely to be solved later.

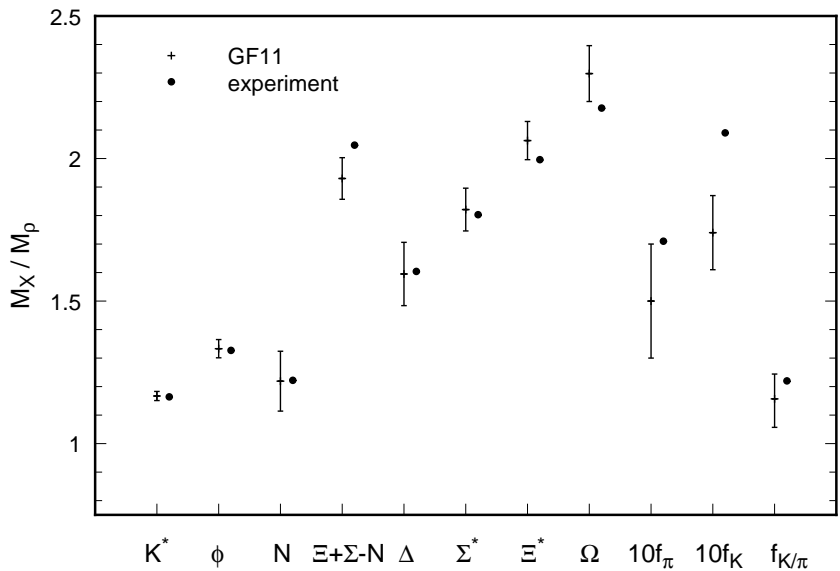

Figure 4: The results from Ref. [41] in comparison with experiment

Nevertheless, I would like to discuss some of the progress of the last few years [40], because it has been significant. The results of a systematic calculation of the light hadron spectrum in the quenched approximation [41] are shown in Figure 4. The inverse lattice spacing was determined from the mass of the $\rho$, the pion and kaon masses were used to set the quark masses. All the predicted hadron masses in Figure 4 are written in units of the $\rho$ mass. Three sources of systematic error were corrected by extrapolations in $m_q$ (to $m_{u,d}$), to zero lattice spacing, and to infinite volume. The comparison of the results of this calculation with the experimentally observed spectrum indicates that quenched QCD might not be a bad approximation to QCD. However, some of the assumptions used in the extrapolations still need to be checked. Lattice-Spacing errors can be reduced with little effort by considering an improved action instead of the Wilson action which was used. A similar analysis with the staggered fermion action is in progress [42].

Calculations that include sea quarks have been improved significantly over the last few years. However, with the exception of finite-volume effects [43], systematic errors ($a \to 0$, $m_q \to m_{u,d}$) have not been extrapolated away yet. On the other hand, the gauge configurations generated with sea quarks are useful for quarkonium physics, because lattice-spacing errors can be controlled and light-quark mass effects are most likely mild.

The effects of the quenched approximation have also been studied in chiral perturbation theory [44]. This serves to guide our expectations about the limitations of the quenched approximation and to identify physical quantities that are insensitive to sea-quark effects.

In principle, numerical calculations of the light hadron spectrum together with perturbation theory can lead to determinations of the light-quark masses. While this should be very interesting in the future, since the light-quark masses are very poorly known, the systematic errors at present make such determinations unreliable [45].

# 5  The CKM matrix

The CKM matrix is a window to new physics, because it is linked to the unknown fermion mass generation mechanism. An overconstrained CKM matrix can give us hints about the physics beyond the Standard Model such as the existence of a fourth generation of quarks and leptons or sources of CP violation inconsistent with the one allowed phase in the unitary matrix. Information about CKM matrix elements can, in principle, be extracted from experimental measurements of exclusive weak processes, such as neutral meson mixing, semi-leptonic and leptonic decays, and flavor-changing radiative decays. However, these processes also receive QCD corrections, the long distance part of which is parametrized into a weak matrix element (see section 2.3, Eq. (11)). With the exception of a few special cases, these matrix elements cannot be calculated from first principles using analytic methods. Lattice QCD can therefore play an important role for this phenomenology (see also recent reviews [46]).

## 5.1  $K - \bar{K}$ Mixing

In the Standard Model $K - \bar{K}$ mixing proceeds through the exchange of two $W$ bosons [47]:

$$\varepsilon \sim \mathrm{Im}(V_{td})|V_{cb}|[(f_3(m_t)\eta_3 - \eta_1 y_c)|V_{cd}| + \eta_2 y_t f_2(m_t)\mathrm{Re}(V_{td})]\frac{3}{8}m_K^2 f_K^2 \boldsymbol{B_K}  , \qquad (22)$$

$$B_K = \frac{3 \langle \bar{K}|\mathcal{O}_{\Delta S=2}|K\rangle}{8 m_K^2 f_K^2} . \qquad (23)$$

The $f_i$, $y_i$ are known functions of the (charm and top) quark masses and $\eta_i$ are perturbative QCD corrections. Using three-generation unitarity, the dependence of $\varepsilon$ on the CKM angles can be reduced to the CP violating phase and $|V_{ub}/V_{cb}|$. $B_K$ is for a variety of reasons one of the simplest quantities to study in lattice QCD. Because it is the ratio of two very similar matrix elements, many systematic errors are expected to cancel. The chiral symmetry of staggered fermions makes them the method of choice for $B_K$ [48].

Most systematic effects have been studied for this quantity. While all groups see a sizable lattice-spacing dependence in their results [49, 50], this error has recently been shown to be $\mathcal{O}(a^2)$ on theoretical grounds [51], leading to a significant reduction of the extrapolation errors. The expectation that errors associated with the quenched approximation cancel has been checked at somewhat coarse lattice spacings with sea quarks of mass $m_q \simeq m_s/2$ [50, 52]. The perturbative corrections to several different operators have been calculated [53, 54]. Their inclusion yields consistency between those operators after extrapolation to zero lattice spacing [50, 51]. The final result in the naive dimensional reduction scheme is [51]:

$$B_K(NDR, 2\,\text{GeV}) = 0.616 \pm 0.020 \pm 0.017 \quad , \tag{24}$$

which can, of course, be converted to the more customary scale independent parameter by $\hat{B}_K = \alpha_s^{-6/25} B_K(NDR, 2\,\text{GeV})$. The first error in Eq. (24) is statistical, the second is the sum of all systematic errors considered. Still to be checked is the lattice-spacing dependence of calculations with $n_f \neq 0$, which need not equal the $n_f = 0$ case. While sea quarks with non-degenerate masses, $m_q \neq m_s$, still need to be incorporated into numerical simulations, they have been estimated [44] in chiral perturbation theory to affect $B_K$ at the less than 5% level. In summary, Eq. (24) is the result of a careful analysis, with most errors under control. It already surpasses all other (model dependent) determinations of this quantity in accuracy. Once the last bit of "hand waving" with respect to sea-quark effects is checked directly, this result will significantly reduce the uncertainty in the Wolfenstein parameters $\rho$ and $\eta$.

## 5.2 $B - \bar{B}$ Mixing

This case is very similar to $K - \bar{K}$ mixing. The amplitude is written in the form:

$$x_d \sim |V_{td}^* V_{tb}|^2 f(m_t) \frac{8}{3} m_B^2 \boldsymbol{f}_B^2 \boldsymbol{B}_B \quad . \tag{25}$$

The major uncertainty in extracting the CKM angles from the experimental measurement of $x_d$ comes from the unknown decay constant $f_B$. $B_B$ is expected to be close to the vacuum saturation value, $B_B \simeq 1$ [55].

Heavy meson decay constants have been studied in various limits by a large number of groups. A lot of the simplifying features that allowed easy control over systematic errors for $B_K$ are absent here. Hand waving arguments about cancellations of systematic errors can only be made for ratios of decay constants, $f_B/f_D$, $f_{B_s}/f_B$, $f_{D_s}/f_D$, which have also been considered.

The $B$ meson in the static limit [56] is an example of a quantity where (for well understood reasons) the extraction of the ground-state signal from the correlator is difficult. This systematic error has been the focus of two groups [57, 58]. They use sophisticated techniques of removing excited states by constructing a matrix of hadron propagators over several different operators. Attention has also been paid to finite-volume and lattice spacing effects [57, 59, 60]. The results for $f_B$ in the static limit are summarized in Figure 5. Some of the disagreement follows from the difficulty of extracting the ground-state signal, which has not been addressed by all groups in the same detail. The lattice-spacing dependence in Figure 5 is rather strong and needs to be better understood or reduced, before the $a \to 0$ extrapolation becomes reliable. This dependence will be somewhat softened by the inclusion of perturbative corrections [61], which have been left out for the purpose of comparison. Also, using an improved action instead of the Wilson action for the light quarks should reduce the lattice-spacing errors.

The decay $D_s \to \mu\nu$ has been measured in several experiments [62], yielding determinations of $f_{D_s}$ (assuming $V_{cs}$). Lattice calculations of this quantity can then be used to calibrate the quenched approximation once all other systematic errors have been removed and the experimental results become more precise. Figure 6 shows a comparison of lattice and experimental results. Systematic errors were considered to varying degrees and are not yet completely controlled (even within the quenched approximation). One group [66] has partially included sea-quark effects.

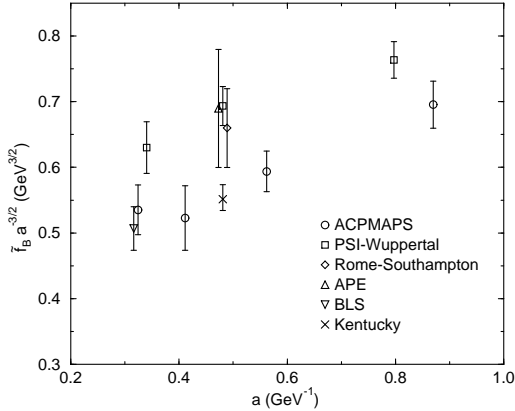

Figure 5: Lattice QCD results for $f_B^{\text{stat}}$ in comparison. ○: [57]; □: [60]; ◇: [64]; △: [63]; ▽: [59]; ×: [58].

For the $B$ meson decay constant, results for mesons containing one heavy quark of order the charm quark (or heavier) mass have traditionally been combined with the static limit, and interpolated to the $B$ meson mass. In the heavy-quark limit the combination $\phi = f\sqrt{m}$ becomes independent of the heavy-quark mass [68] (up to renormalization group logarithms), with corrections that vanish as powers of $1/m$:

$$\phi = \phi_\infty \left(1 + \frac{c_1}{m} + \frac{c_2}{m^2} + \ldots\right) \qquad (26)$$

An example of this interpolation from Ref. [59] is shown in Figure 7; similar results were obtained by all groups. In particular, large deviations from the static limit as shown in Figure 7 were seen by all groups. However, lattice-spacing errors in the heavy-quark action (and current) also contribute $1/m$ terms. Even though they are probably small, they should be removed by using improved heavy-quark actions and currents [10, 14]. Improved heavy-quark methods can also obviate the (expensive) need to use ever smaller lattice spacings for weak matrix elements evaluated directly at the $B$ meson mass.

I conclude this section with a summary of the current state of knowledge of heavy meson decay constants in Table 1. I would like to emphasize again that, while attention has been paid to systematic errors, they are not yet under complete control in the quenched approximation. However, there are no technical obstacles to calculations of heavy meson decay constants in the continuum limit of quenched lattice QCD. While the inclusion of sea-quark effects will be more difficult to control, ratios of decay constants may prove to be as insensitive to these and other systematic errors as $B_K$ has been shown to be.

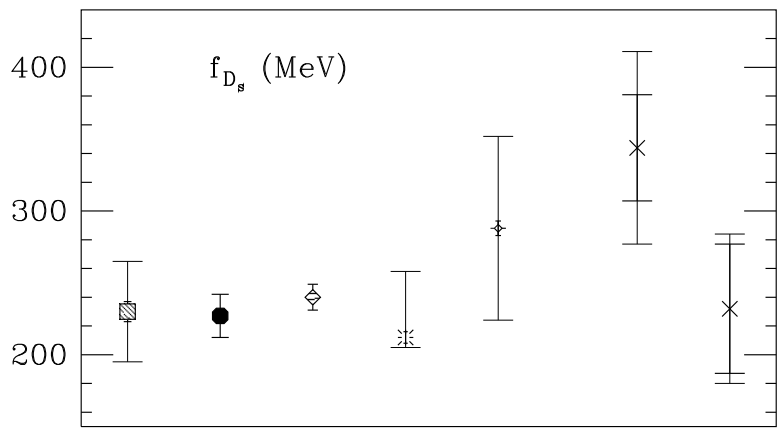

Figure 6: A Comparison of lattice QCD and experimental results for $f_{D_s}$. The second error is the groups estimate of the systematic errors they considered. ×: Experiment [62]; □: BLS [59]; •: ELC [64]; ⋄: APE [63]; ∗: UKQCD [65]; ⋄: HEMCGC [66].

| group | $f_D$ (MeV) | $f_{D_s}/f_D$ | $f_B$ (MeV) | $f_{B_s}/f_B$ |
|---|---|---|---|---|
| BLS [59] | $208 \pm 9 \pm 37$ | $1.11 \pm 0.02 \pm 0.05$ | $187 \pm 10 \pm 37$ | $1.11 \pm 0.02 \pm 0.05$ |
| UKQCD [65] | $185 \pm^4_3 \pm^{42}_7$ | $1.18 \pm 0.02$ | $160 \pm 6 \pm^{53}_{19}$ | $1.22 \pm^4_3$ |
| PWCD [60] | $170 \pm 30$ | $1.09 \pm 0.02 \pm 0.05$ | $180 \pm 50$ | $1.11 \pm 0.02 \pm 0.05$ |
| APE [63] | $218 \pm 9$ | $1.11 \pm 0.01$ | | |
| ELC [64] | $210 \pm 15$ | $1.08 \pm 0.02$ | $205 \pm 40$ | $1.08 \pm 0.06$ |
| HEMCGC [66] | $215 \pm 9 \pm 53$ | | | |
| FNAL [57] (static) | | | $188 \pm 23 \pm^{34}_{21}$ | $1.216 \pm 0.041 \pm 0.016$ |
| SH [67] (non-rel.) | | | $171 \pm 22 \pm^{19}_{45}$ | |

Table 1: Lattice QCD results for heavy mesons decay constants in comparison. The second error is the groups estimate of the systematic errors they considered.

### 5.3 Semi-Leptonic Decays

Semi-leptonic decays have traditionally been used to determine CKM matrix elements. The two classic examples are nuclear $\beta$-decay for $V_{ud}$ and $K_{l3}$ for $V_{us}$. In both cases, ($SU(2)$ or $SU(3)$) chiral symmetry together with current conservation was used to calculate the hadronic matrix element. Semi-leptonic decays of charm or bottom mesons have no such advantage (with the exception of decays like $B \to D^{(*)} l \nu$, see section 5.4.1). Hence, in all but this one case exclusive semi-leptonic decays of $D$ and $B$ mesons have failed to provide accurate determinations of the corresponding CKM matrix elements, even though a wealth of experimental information has been accumulated in recent years [69].

The matrix element for, say, $D \to K^{(*)} l\nu$ can be parametrized in terms of form factors. Common parametrizations are:

$$\langle K|J_\mu|D\rangle = f_+(q^2)\left[(p_D+p_K)_\mu - \frac{m_D^2-m_K^2}{q^2}(p_D-p_K)_\mu\right] + f_0(q^2)\frac{m_D^2-m_K^2}{q^2}(p_D-p_K)_\mu \quad (27)$$

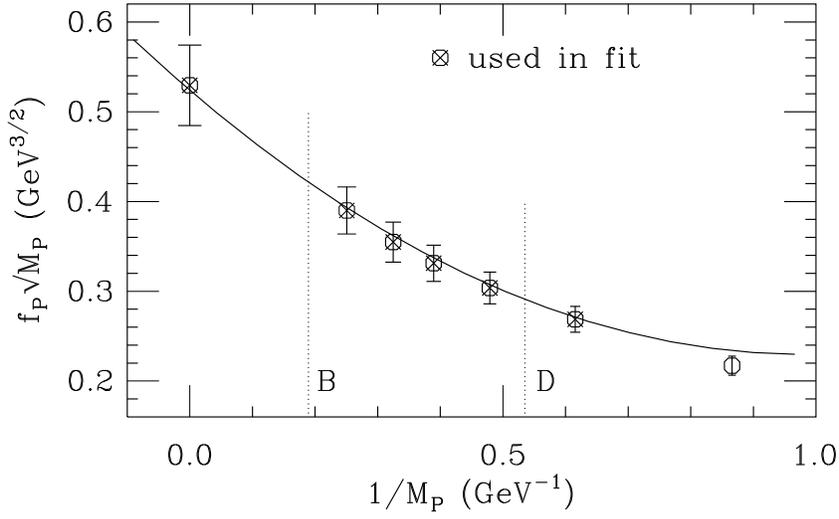

Figure 7: $\phi$ vs. $1/m$ from Ref. [59].

and

$$\langle K^*, \lambda | J_\mu | D \rangle = \epsilon_\alpha^{(\lambda)} \left[ \frac{2}{m_D + m_{K^*}} V(q^2) \epsilon_{\mu\alpha\rho\sigma} p_D^\rho p_{K^*}^\sigma - A_1(q^2)(m_D + m_{K^*})\delta_{\mu\alpha} \right. \tag{28}$$
$$\left. + A_2(q^2) \frac{1}{m_D + m_{K^*}} (p_D + p_{K^*})_\mu p_{D\alpha} - A(q^2) \frac{2m_{K^*}}{q^2} (p_D - p_{K^*})_\mu p_{D\alpha} \right]$$

with momentum transfer $q^2 = (p_D - p_{K^{(*)}})^2$. In the rest frame of the $D$ meson, zero recoil ($\mathbf{p}_{K^{(*)}} = 0$) corresponds to maximum momentum transfer, $q_{\max}^2 = (m_D - m_{K^{(*)}})^2$. Maximum recoil is at $q^2 = m_l^2 \approx 0$. Knowledge of the form factors yields CKM matrix elements from measurements of, for example, the differential decay rate:

$$\frac{d\Gamma(D \to K l \nu)}{dq^2} = |V_{cs}|^2 \frac{G_F^2 \lambda^{3/2}}{192\pi^3 m_D^3} |\boldsymbol{f}_+(\boldsymbol{q^2})|^2 \quad, \tag{29}$$

or

$$\frac{d\Gamma(D \to K^* l \nu)}{dq^2} = |V_{cs}|^2 \frac{G_F^2 \lambda^{1/2} q^2}{64\pi^3 m_D^3} (m_D + m_{K^*})^2 |\boldsymbol{A}_1(\boldsymbol{q^2})|^2 + \mathcal{O}(\lambda^{3/2}) \tag{30}$$

where $\lambda \equiv \lambda(m_D, m_{K^{(*)}}, q^2) = [m_D^2 - m_{K^{(*)}}^2 - q^2]^2 - 4m_{K^{(*)}}^2 q^2$. It is evident from Eqs. (29) and (30) that the rate vanishes at zero recoil, because of the kinematic factor $\lambda^{1/2} \equiv 2m_D \mathbf{p}_{K^{(*)}}$.

It is therefore necessary to consider the form factors as functions of the recoil momentum (or $q^2$). However, momenta are discretized by the finite volume, and lattice artifacts increase as $a\mathbf{p} \sim 1$. Finite-Volume errors should also be expected to change relative to the zero momentum case. Mesons with finite momentum have similar problems with the time dependence of statistical errors as mesons containing a static quark (see section 5.2) or orbitally excited states, which makes the ground-state extraction difficult. The overall increase in statistical errors relative to the zero momentum case obfuscates systematic error analyses.

All groups limit themselves to momenta with $|\mathbf{p}| \leq 2p_{\min}$ ($p_{\min} = \frac{2\pi}{L}$). The form factors are calculated at the corresponding discrete values of $q^2$. Smart choices of initial and final state momentum combinations can still cover a significant range of momentum transfers.

The CKM angle $V_{cs}$ is known from three-generation unitary of the CKM matrix. The decay $D \to K^{(*)} l \nu$ can therefore be used to test lattice methods and calibrate the quenched approximation. A comparison is shown in Table 2. As pointed out in Ref. [70], another nice calibration mode is the semi-leptonic decay $K \to \pi l \nu$, because precise measurements of both form factors, $f_+$ and $f_0$, as functions of $q^2$ exist.

| group | $f_+(q^2_{\max})$ | $A_1(q^2_{\max})$ | $A_2/A_1(q^2_{\max})$ | $V/A_1(q^2_{\max})$ |
|---|---|---|---|---|
| Exp [69] | $1.31 \pm 0.05$ | $0.66 \pm 0.05$ | $0.73 \pm 0.15$ | $2.04 \pm 0.27$ |
| ELC [71] | $1.13 \pm 0.31$ | $0.62 \pm 0.09$ | $0.70 \pm 0.40$ | $1.62 \pm 0.30$ |
| APE [72] | $1.36 \pm 0.14$ | $0.79 \pm 0.13$ | $0.70 \pm 0.40$ | $1.73 \pm 0.32$ |
| UKQCD [73] | $1.17 \pm ^{0.12}_{0.14}$ | $0.83 \pm ^{0.07}_{0.12}$ | $0.90 \pm 0.20$ | $1.50 \pm ^{0.50}_{0.20}$ |
| LANL [74] | $1.27 \pm 0.10$ | $0.78 \pm 0.04$ | $0.72 \pm 0.22$ | $2.03 \pm 0.13$ |
| BKS [70] | $1.64 \pm 0.36 \pm 0.36$ | $1.23 \pm 0.2 \pm 0.31$ | $0.70 \pm 0.16 \pm ^{0.20}_{0.15}$ | $2.15 \pm 0.24 \pm ^{0.33}_{0.38}$ |

Table 2: Lattice QCD results for $D \to K^{(*)} l \nu$ in comparison with experiment. The second error is the groups estimate of the systematic errors they considered.

For historical reasons, experimentalists and (lattice) theorists alike have quoted the form factors at $q^2 = 0$ (which is usually coupled with parametrizing the $q^2$ dependence by pole dominance), even though this kinematic point is not particularly convenient for either group. Eqs. (29) and (30) emphasize that experimental measurements of differential decay rates and lattice determinations of form factors at the same recoil momenta can be used to extract the CKM angles without model assumptions about the $q^2$ dependence. If such assumptions have to be made because of the experimental set-up, I suggest to quote the form factors at $q^2 = q^2_{\max}$ for comparison with theory. To emphasize this point, I have changed the standard comparison of form factors at $q^2 = 0$ to $q^2 = q^2_{\max}$ in Table 2. The experimental results were simply rescaled by the appropriate factor ($[1 - q^2_{\max}/m^2_{\text{pole}}]^{-1}$); for the lattice QCD results I took the form factors as quoted at $q^2_{\max}$ where available, otherwise they were rescaled in the same fashion as the experimental results.

A study of lattice-spacing, finite-volume, and other systematic errors in Ref. [70] was very limited by large statistical uncertainties. While several groups have presented calculations since [71, 72, 73, 74], no systematic investigation of finite-volume and finite lattice-spacing effects has appeared yet.

Most groups [70, 71, 72] have also considered other charm decays, like $D \to \pi(\rho) l \nu$, and $D_s \to \eta(\phi) l \nu$, where possible contributions from non-spectator diagrams were neglected.

Ref. [71, 72] extrapolated their results for charm decays to $B \to \pi(\rho) l \nu$ assuming heavy quark symmetry. An analysis of $B$ meson semi-leptonic decays at realistic $b$ quark masses has not been carried out yet.

In summary, the need to consider mesons with nonzero momentum complicates lattice QCD calculations of semi-leptonic decays relative to calculations of decay constants. So far, lattice calculations of semi-leptonic charm decays have been less systematic than those of heavy meson decay constants. However, no serious obstacles exist, to obtaining definitive results in the quenched approximation at least over a limited range of recoil momenta.

## 5.4  Other Weak Matrix Elements

In the following two subsections I discuss the status of calculations of the Isgur-Wise function (relevant for experimental measurements of $B \to D^{(*)} l \nu$) and of the exclusive radiative decay $B \to K^* \gamma$.

In both cases the first numerical lattice QCD results appeared only very recently. The lattice techniques are, of course, very similar to those established for the study of semi-leptonic charm decays.

### 5.4.1 The Isgur-Wise Function

Experimental measurements of the differential decay rates for the exclusive decays $B \to D^* l \nu$ near zero recoil depend with only small corrections on the Isgur-Wise (IW) function. Using Heavy Quark Effective Theory (HQET) it has been shown that the corrections to the symmetry limit are small for this mode, yielding a model-independent determination of $V_{cb}$ [68, 69]. For present day lattice calculations this mode is useful as a calibration of the quenched approximation.

Two different approaches to lattice QCD calculations of the IW function [75] have been employed, in analogy to the $B$ meson decay constant. The discretization of the Heavy Quark Effective Theory allows for the direct computation of the IW function [76] from the corresponding matrix elements.

The IW function can be extracted from matrix elements which are related to the above process by HQET. In particular, the corrections to heavy quark symmetry for the form factors extracted from matrix elements of the form $\langle D|\bar{c}\gamma_\mu c|D\rangle$, which were the focus of the initial lattice QCD calculations [77, 78], are similarly suppressed. Ref. [78] also considered the processes "$B$" $\to D^{(*)} l \nu$ using heavy quark masses around charm. Figure 8 shows as an example a comparison of the results of Ref. [78] with experimental data from CLEO [79].

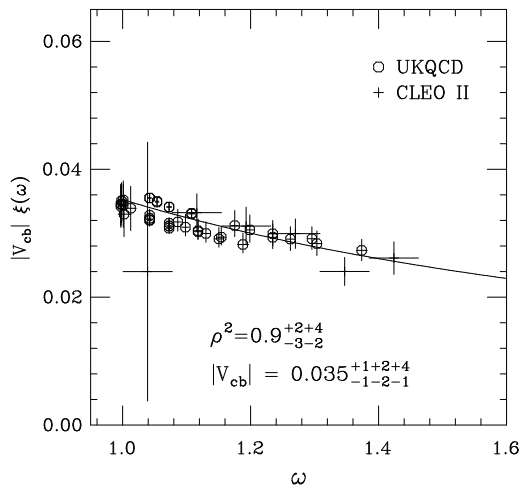

Figure 8: The IW function from Ref. [73] in comparison with CLEO II data.

The knowledge of systematic errors is similar to the case of semi-leptonic charm decays. No serious analysis of finite volume, finite lattice spacing and (heavy and light) quark mass effects exists yet.

The error on $V_{cb}$ can, in principle, be reduced beyond the current knowledge, once lattice QCD calculations of the form factors at recoil momenta where the differential decay rate is measured are at least as accurate as the current theoretical estimate at zero recoil, because the at present necessary extrapolation of experimental results to zero recoil can then be avoided.

### 5.4.2 $B \to K^* \gamma$

Lattice calculations of the radiative decay, $B \to K^* \gamma$, have also recently begun [80, 81], where, as before, heavy quark masses of order the charm quark were used. The exclusive decay rate is calculated with the hadronic matrix element $\langle K^*|\bar{s}\sigma_{\mu\nu}q^\nu b_R|B\rangle$, which is again parametrized in terms of form factors, $T_1$ and $T_2$. The physical two-body decay takes place at momentum

transfer $q^2 = 0$. Extrapolations to $m_B$ using HQET and to $q^2 = 0$ using pole dominance are therefore necessary. Some consistency checks are possible by interchanging the order of the extrapolations.

The ratio
$$R_{K^*} = \frac{\Gamma(B \to K^*\gamma)}{\Gamma(b \to s\gamma)} = 4[m_B/m_b]^3[1 - \frac{m_{K^*}^2}{m_B^2}]|\boldsymbol{T}_1(\mathbf{o})|^2 \tag{31}$$
has been measured [82] to $R_{K^*} = (19 \pm 8)\%$. Lattice QCD predictions of this ratio [80, 81] are in the range $R_{K^*} = 6 - 9\%$ with an uncertainty of about 50 %. Given the exploratory nature of the lattice QCD calculations and the large experimental errors, this is pretty fair agreement.

These results should be followed up by calculations directly at the $B$ meson mass, together with a systematic study of lattice-spacing artifacts at large recoil.

## 6 Conclusions

Lattice QCD is becoming a reliable tool for Standard Model phenomenology, even though we are far from being done yet.

With present day technology, it will be challenging to reach the level of control over systematic errors in calculations that include sea-quark effects that is possible in calculations that use the quenched approximation. This approximation will therefore be with us for some time. Definitive calculations of the light hadron spectrum, heavy meson decay constants and other simple weak matrix elements in the continuum and infinite-volume limit of quenched lattice QCD are possible. We will continue to see such results in the coming years. The comparison with the experimentally measured light hadron spectrum and weak decays (where the corresponding CKM matrix elements are known) can give us a calibration of quenched QCD. While the final judgement on whether quenched QCD can give us a useful phenomenology is still pending, most results in quenched QCD so far indicate that it will.

In the meantime, progress will continue to be made in calculations including sea-quark effects, leading to phenomenologically interesting results and in due course to better knowledge of an increasing number of Standard Model parameters. The search for better lattice actions (with very small discretization errors at, say, lattice spacings of $a \sim 0.5$ fm) and better algorithms (for the inclusion of sea quarks) is a very active area of research – I invite the reader to take a look at the proceedings of the 1994 meeting on Lattice Field Theory [83]. This has, of course, the possibility of revolutionizing the field, making possible definitive results sooner than we anticipate.

In some special cases, like quarkonium spectroscopy and $B_K$, such calculations are already possible with current technology and with control over systematic errors. A first-principles understanding of these (and hopefully other) simple physical systems should be possible within the next few years.

## Acknowledgements

I thank the organizers for an enjoyable conference and the authors of the work reviewed here for correspondence while preparing this talk. Discussions with C. Bernard, C. Davies, E. Eichten, S. Gottlieb, A. Kronfeld, P. Lepage, P. Mackenzie, and J. Shigemitsu are also gratefully acknowledged.